\documentclass [prc,preprint,showpacs]{revtex4}

\usepackage{graphicx}
\begin{document}
\title{Isospin equilibration processes and dynamical correlations:
study of the system $^{40}Cl+^{28}Si$ at 40 MeV/nucleon}
\author{M.Papa$^{a)}$\footnote{e-mail: papa@ct.infn.it} and G.Giuliani $^{b)}$}
\affiliation{\textit{ a) Istituto Nazionale Fisica Nucleare-Sezione
di Catania, V. S.Sofia 64 95123 Catania Italy}}
\affiliation{\textit{
 b)Dipartimento di Fisica e Astronomia, Universit\'a di Catania
 V. S.Sofia 64 95123 Catania Italy}}
\begin{abstract}
The asymptotic time derivative of the total dipole signal is
proposed as an useful observable to investigate on Isospin equilibration
phenomenon in multi-fragmentation processes.
The study
has been developed to describe charge/mass equilibration processes
involving the gas and liquid "phases" of the total system formed
during the early stage of a collision. General properties of this observable
and the links with others isospin dependent phenomena are
discussed.
In particular, the $^{40}Cl+^{28}Si$ system at 40
MeV/nucleon is investigated by means of semiclassical microscopic many-body
calculations based on the CoMD-II model.  The study of the dynamical many-body correlations produced by the model
 also shows how the proposed observable is rather sensitive to different parameterizations of the
isospin dependent interaction.

\end{abstract} 
\maketitle
\section{Introduction}
An interesting subject related to Heavy Ions Isospin physics
\cite{schr} is the process  leading to the equilibration of the
charge/mass ratio between the main partners of the reaction as well
described in Ref.\cite{pawel}. The so-called "isospin diffusion"
phenomenon is the relevant mechanism acting between the reaction
partners in binary processes \cite{bettyprl,betty,baodif,dtoro,gal}. In
particular, in  the collision of the 124 and 112 Tin isotopes at 50
MeV/nucleon \cite{bettyprl}, evidence of partial equilibrium in the
charge/mass ratios of the quasi-projectile and quasi-target has been
deduced through the study of the iso-scaling parameters related to
the isotopic distributions. In this case dynamical
calculations, based on the Boltzmann-Uehling-Uhlenbeck
model \cite{buu,betty1}, show that the degree of equilibration
depends on the behavior of the symmetry potential $U^{\tau}$ as a
function of the density.

\noindent
In this work we want to extend the study of the isospin
equilibration processes, looking at the system in a global way, by using the
following quantity:
$\overrightarrow{V}(t)=\sum_{i=1}^{Z_{tot}}\overrightarrow{v}_{i}$.
 At a microscopic level, the sum on the index
$i$ is performed on all the $Z_{tot}$ protons (bound and free) of the system.
$\overrightarrow{V}(t)$ corresponds, apart from the elementary
charge $e$, to the time derivative of the total dipole of the
system. The velocities $\overrightarrow{v_{i}}$ are computed in the
center of mass (c.m.) reference frame. We note that, as due to the
total momentum conservation, the global effect related to the motion
of the neutral particles (bound and free) is also implicitly contained
in $\overrightarrow{V}(t)$. Therefore we can expect, in a quite general
way, a peculiar dependence of the behavior of $\overrightarrow{V}(t)$
from the iso-vectorial interaction.

Several studies were based on this dynamical variable to describe
pre-equilibrium Giant Dipole Resonance (GDR) $\gamma$-ray emission (see
Ref.\cite{asygdr,ca10mev,trasex1} and references therein ). Various
reasons suggest us to use the same variable to also describe isospin
equilibration in complex processes.
\noindent
The basic
guide line  starts just from these studies on pre-equilibrium radiative emission
in essentially binary processes induced by
charge/mass asymmetric partners.  The ensemble average of the time derivative of the
total dipole, in the initial and final stages of
the collision, can be approximated through  the so-called Molecular component \cite{12C}:

\begin{eqnarray}
\langle\overrightarrow{V}\rangle & \approx &
\langle\overrightarrow{V}_{M}\rangle\
\equiv\frac{1}{2}\langle\mu_{PT}\rangle(\langle \beta_{T}\rangle-\langle \beta_{P}\rangle)
\langle\overrightarrow{v_{PT}}\rangle
\end{eqnarray}
 where $\mu_{PT}$ is the reduced
mass of the projectile or quasi-projectile (P) and target or
quasi-target (T) of the  binary system,
$\beta_{P,T}=\frac{N_{P,T}-Z_{P,T}}{A_{P,T}}$ represents the
associated relative neutron excess ($N_{P,T},Z_{P,T},A_{P,T}$ are
the neutron, proton and mass numbers respectively),
$\overrightarrow{v_{PT}}$ indicates the relative velocity.
During the interaction stage of the collision, if
the partners have exchanged charge and mass for enough time in such a way
to equilibrate, on average, the charge/mass ratio, then the final value
of $\langle \overrightarrow{V} \rangle$ will be zero. In this case the yield of the
pre-equilibrium dipolar $\gamma$-ray emission, which  satisfies the well known selection rule
on the isospin quantum number $T$ ($\Delta T=1$, no isospin mixing)
of the involved intermediate states, reaches the maximum value.
For more complex mechanisms, as the ones leading to the substantial
stopping of the two partners and to the formation of one hot source,
the dipolar signal related to the
 other particles and fragments can not be neglected.
 However, also in these cases we can use the
same definition of isospin equilibration through the condition
$\langle\overrightarrow{V}\rangle=0$. As we will show in the
following, this definition leads to other conditions concerning the
production of differential flow between neutral and charged nucleons
\cite{bali00,bali08} and/or the relative neutron excesses of the
main produced fragments and free particles. \vskip 5pt \noindent
 In the following we briefly discuss some simple
examples and properties of $\langle\overrightarrow{V}\rangle=0$
 aiming
to illustrate the information which is potentially contained in
this quantity and the relation with charge/mass equilibration
processes:

 - (i) By taking into account  only the effects associated to the strong interaction, after the pre-equilibrium
 stage, starting from the time $t_{pre}$, when a second stage characterized by
 an average isotropic  or symmetric emission of the secondary sources (statistical equilibrium)
 takes place, the ensemble average of $\overrightarrow{V}(t)$
 satisfies the following relation:
 $\langle\overrightarrow{V(t_{pre})}\rangle=
 \langle\overrightarrow{V(t>t_{pre})}\rangle\equiv\langle\overrightarrow{V}\rangle$ \cite{ca10mev}.
The average value of this dynamical variable at $t_{pre}$
 is invariant with respect to statistical processes
 and therefore the value of $\langle\overrightarrow{V}\rangle$ is determined
only by the complex dynamics which characterizes the early stage of
the collision, when fast changes of the average nuclear density are
expected. In particular, $\langle\overrightarrow{V}\rangle$
 can be expressed as a function of the charge $Z$, mass $A$, average multiplicity
 $\langle {m}_{Z,A} \rangle$ and the average value of the mean momentum
 $\langle \overrightarrow{P}_{Z,A}\rangle$ per unit of mass (expressed in $fm^{-1}$)
 of the detected particles  in the generic event:
\begin{eqnarray}
\langle \overrightarrow{V} \rangle =\sum_{Z,A}\frac{Z}{A}\langle
m_{Z,A} \rangle \langle \overrightarrow{P}_{Z,A} \rangle
C_{\overrightarrow{P}}^{Z,A}\\
C_{\overrightarrow{P}}^{Z,A}= \frac{\langle m_{Z,A}
\overrightarrow{P}_{Z,A} \rangle} {\langle
\overrightarrow{P}_{Z,A}\rangle \langle m_{Z,A}\rangle}
\end{eqnarray}

$C_{\overrightarrow{P}}^{Z,A}$ is the correlation function between
the multiplicity and the mean momentum. This correlation function
plays a key role for the invariance property and therefore requires
for an  event by event analysis in which many-body correlations can
not be neglected. For symmetry reasons, $\langle \overrightarrow{V}
\rangle$ lies on the reaction plane. It is directly linked with a
weighted mean of the charge/mass ratio, as Eq.(3) suggests. It also
takes into account the average isospin flow direction through the
momenta $\overrightarrow{P}_{Z,A}$.
\noindent
The long range Coulomb interaction  can produce
differences between the value of
$\langle\overrightarrow{V(t_{pre})}\rangle$ and the observed
asymptotic value. These changes, however, are rather small at the
involved energies (see Sec.III) and  can be evaluated with the necessary precision by
taking into account the corrections due to the Coulomb repulsion up
to the asymptotic stage.

 -(ii) In the general case, we
find attractive the following decomposition:
$ \langle \overrightarrow{V} \rangle = \langle \overrightarrow{V}_{G} \rangle +
\langle \overrightarrow{V}_{L} \rangle + \langle \overrightarrow{V}_{GL} \rangle$
where $ \langle \overrightarrow{V}_{G} \rangle$ and
$ \langle \overrightarrow{V}_{L} \rangle $ are the average dipolar signals
associated to the gas "phase" (light charged particles) and to the
"liquid" part \cite{indra} corresponding to the motion of the
produced heavy fragments. The signal
$ \langle \overrightarrow{V}_{GL} \rangle$ is instead associated to the
relative motion of the two "phases". By supposing, for simplicity,
the gas "phase" formed by neutrons and protons,
$ \langle \overrightarrow{V} \rangle$ can be further decomposed as:
\begin{eqnarray}
\langle \overrightarrow{V} \rangle =
\langle\frac{A_{G}(1-\beta^{2}_{G})}{4}\overrightarrow{v}_{r}^{PN} \rangle+
\langle\frac{\mu_{GL}(\beta_{L}-\beta_{G})}{2}\overrightarrow{v}_{cm,LG}\rangle+
\langle\overrightarrow{V}_{r,L}\rangle
\end{eqnarray}
In the above expression the first term represents the contribution
related to the proton-neutron relative motion of the gas "phase"
composed of $A_{G}$ nucleons, expressed through the relative
velocity $\overrightarrow{v}_{r}^{PN}$; the second term concerns the
relative velocity $\overrightarrow{v}_{cm,LG}$ between the centers
of mass of the "liquid" complex and the "gas",  $\mu_{GL}$ is the
reduced mass associate to the two sub-systems; the last term
represents the contribution produced by the relative motion of the
fragments. A similar expression can be obtained including other
light particles in the gas "phase" as , for example, light
Intermediate Mass Fragments (IMF).
 From this decomposition we can see how the
isospin equilibration condition ($\langle
\overrightarrow{V}\rangle=0 $), for the total system, requires a
delicate balance which depends on the average neutron excess of the
produced "liquid drops" $\langle \beta_{L} \rangle$, on the one
associated to the gas "phase" $\langle \beta_{G} \rangle$, and on
the relative velocities
 between the different parts. To enlighten the role
 played by some of the terms reported in Eq.(4),
 we can discuss the idealized decay of a
charge/mass asymmetric source through neutrons and protons emission
(or the case in which the liquid drops are produced through a
statistical mechanism $\langle \overrightarrow{V}_{r,L} \rangle=0$).
This example can also schematically describe the case of the
complete stopping in heavy ion collisions.
 Moreover, for simplicity, we can
consider uncorrelated fluctuations between the velocities, masses
and neutron excesses. In absence of pre-equilibrium emission or for
identical colliding nuclei
 the second term of
Eq.(4) is zero ($\langle \overrightarrow{v}_{cm,LG}\rangle= 0 $),
and the isospin equilibration requires a neutron or proton gas
"phase" or absence of relative neutron-proton motion. For
non-identical colliding nuclei, if pre-equilibrium emission exists,
then $\langle \overrightarrow{v}_{cm,LG}\rangle\neq 0$. In this
case, if $\langle \beta_{G} \rangle \neq \langle \beta_{L} \rangle$,
due, for example, to the isospin "distillation" phenomenon
, the first term has to be necessarily different from zero
and it will contribute to the neutron-proton differential flow (see
also Sect.III). We remark that both the isospin "distillation"
phenomenon \cite{riz} and the production of neutral-charged particles
differential flow \cite{bali08} depend in a peculiar way from the iso-vectorial
forces.

\noindent
Therefore, according to our description, the understanding of the
isospin equilibration process for the total system requires the gas
"phase" contribution to be taken into account. This term can be
regarded as a kind of "dissipation" with respect to the system
formed by the liquid part. In this work, as an example, we will
discuss the results obtained through the Constrained Molecular
Dynamics-II approach (CoMD-II) \cite{comd1,comdII} applied to the
charge/mass asymmetric  $^{40}Cl+^{28}Si$ system at 40 MeV/nucleon.
The study is performed by using different options for the iso-vectorial
potential term. \vskip 5pt \noindent
 Before to show
the results of our calculations, in the following section we briefly
recall the way in which the isospin dependence of the nuclear
interaction is introduced in the CoMD-II model.

\section{Symmetry interaction and correlations}

According to the results shown in Ref.\cite{comd1,epjnew}, starting
from a Skyrme type two-body microscopic interaction, the two-body
effective potential in CoMD-II model can be expressed through the
nucleon-nucleon overlap integral $\rho^{i,j}= \int\int
d^{3}r_{j}d^{3}r_{i}\delta(\overrightarrow{r_{i}}-\overrightarrow{r_{j}})
\widehat{\rho^{i}}\widehat{\rho^{j}}$. $\overrightarrow{r_{j}},\overrightarrow{r_{i}}$ represent
the nucleon spatial coordinates. $\widehat{\rho^{i}}$ is the
Gaussian distribution in the coordinate space related to the generic
nucleonic wave-packet. The microscopic
 iso-vectorial interaction for the Stiff2 option is given by the following expression:
$
V^{\tau} = \frac{a_{0}}{2\rho_{0}}\sum_{i\neq
j=1}^{A}(2\delta_{\tau_{i},\tau_{j}}-1)
  \delta(\overrightarrow{r}_{i}-\overrightarrow{r}_{j})
$, A is the total mass number, $\tau_{i}$ represent the generic
third nucleonic isospin component and $\rho_{0}$ is the one-body density
at the saturation point.
The coefficient $a_{0}=72 MeV$
determines the strength of the iso-vectorial interaction at the
saturation density.
As shown in Ref.\cite{epjnew} the structure of $V^{\tau}$ can be obtained by taking into account
that the two-body nuclear forces, in $S$ wave, around the ground state ($g.s.$) density,  are less attractive in
isospin triplet states ($T=1$) with respect the singlet states ($T=0$).
From the above expressions it results
that the associated effective interaction $U^{\tau}$ (after the convolution with
the nucleonic wave-packets) can be expressed
as a function of the average overlap integrals per couple of
neutrons ($nn$), $\hat{\rho}^{nn}$, protons ($pp$) $\hat{\rho}^{pp}$, and
neutron-proton ($np$) $\hat{\rho}^{np}$. As discussed in
Ref.\cite{epjnew} for small asymmetries we can assume
$\tilde{\rho}^{nn}\cong\tilde{\rho}^{pp}
\cong\frac{\tilde{\rho}^{nn}+\tilde{\rho}^{pp}}{2}=\tilde{\rho}$. To
characterize the differences associated to the nucleon-nucleon
dynamics, at a two-body level,  we can introduce the correlation
coefficient $\alpha$ in such a way $\tilde{\rho}^{np} =
(1+\alpha)\tilde{\rho}$. $\alpha$ depends on both $\tilde{\rho}$ and
the asymmetry parameter $\beta$.
 Results on nuclear matter
simulations \cite{epjnew} show that the behavior of $\alpha$ as a function of
$\beta$ can be approximated for moderate asymmetries by a parabolic
law. In this case, eqs.(6,7) of Ref.\cite{epjnew} give the following
expression for the effective iso - vectorial potential in the  Non
Local (N.L.) approximation $U^{\tau}_{N.L.}$

\begin{eqnarray}
U^{\tau}_{N.L.} & \cong
&\frac{a_{0}}{2\rho_{0}}\hat{\rho}A^{2}F'(s)[(1+\frac{1}{2}\alpha_{0}
-\alpha')\beta^{2}-\frac{1}{2}\alpha_{0}] \\
  \alpha'& = & \frac{1}{4}\frac{\partial^{2} \alpha}{\partial
\beta^{2}}|_{\beta=0}
\end{eqnarray}
 $\beta^{4}$ terms are neglected in the previous expression.
$\alpha_{0}\equiv\alpha(\hat{\rho},\beta=0)$ represents the
correlation coefficient related to the difference in the dynamics of
the $np$ couples with respect to the $nn$ and $pp$ ones for symmetric
nuclear matter.

 It depends on the average overlap integral per couple
of nucleons $\hat{\rho}$ which reflects the degree of compression.
$s=\frac{4}{3A}\sum_{i\neq j}\rho^{i,j}$ is associated to the total
overlap integral per nucleon.  F' is a form factor which modulates the
changes of the iso-vectorial interaction as a function of the
average overlap integral $s$.  For the Stiff1 option we
use $F'=\frac{2s}{s_{g.s.}+s}$, for the Stiff2 case $F'=1$ and for
the Soft option $F'=(\frac{s_{g.s.}}{s})^{1/2}$. These form factors
correspond to the following values of the $K_{sym}$ parameter as defined
in Ref. \cite{bali00, bali08}: 94 MeV for the Stiff1 option, -27 MeV
for the Stiff2 option and -88 MeV for the Soft one. This choice corresponds
also to an $S_{0}$ value \cite{bali00} of about 40 MeV . From Eq.(5) we
note that in our approach the iso-vectorial forces
 generates, beyond the $\beta$
dependent potential, also another iso-vectorial density dependent
term, independent on  $\beta$ and proportional to the
degree of correlation $\alpha$ evaluated for symmetric systems. As discussed in
\cite{epjnew,pyl}, at small asymmetries, this term determines the
high sensitivity of the experimental observables to the different
functional forms of $F'$. The finite value of $\alpha$, which is of
the order of $15\%$, apart from the Pauli principle and the Coulomb
interaction, is strongly affected by the iso-vectorial interaction
itself. In fact  the neutron-proton couples, at
variance with the other ones, suffer the more attractive singlet
interaction.
\vskip 1pt
\noindent
For clarity, we have to note that the $np$ correlations generated by similar kind of forces
represent a rather interesting subject for the description of the extra-bind energies
in $N\simeq Z$ nuclei. These studies are performed with static sophisticated quantal approaches.
The $np$ pairing correlations, for example, play a main role to microscopically describe
the Wigner energy in symmetric nuclei \cite{a,b,c}.
However, apart from effects induced by the Pauli
principle in phase-space (as treated in our approach) and from
the usage  of  an iso-vectorial potential which reflects  a two-body interaction
dependent on  isospin quantum numbers,
the description and the propagation of the many-body correlations
in our model calculations are essentially classical.
The $np$ correlations discussed in the present work  can
represent  a dynamical classical analogous which includes Pauli principle and  isospin effects of more complex
quantal effects.
Therefore
the effects here discussed
can not be compared  with the  ones able to go
beyond this classical analogous and having therefore a pure quantal nature.
In fact
our approach can not describe effects that, for example,
 could specifically arise from  details of the single particles wave functions or also from details  concerning
 the propagation and symmetries of the Slater determinant structure of
the many-body wave functions,
which instead  characterizes
the Fermionic molecular dynamics approaches \cite{feld,ono}.

\noindent
The case corresponding to vanishing values of the
correlation $\alpha$ represents, in our  frame-work, the so-called
Iso-vectorial Mean Field Approximation
(I.M.F.A.).
 In this case the average overlap integrals per couple of nucleons
related to neutron-neutron, proton-proton and neutron-proton
interactions have the same values
($\tilde{\rho}^{nn}=\tilde{\rho}^{pp}=\tilde{\rho}^{np}$) and the
iso-vectorial interactions generate only the usual symmetry
potential term which depends on $\beta^{2}$.

\vskip 5pt
 \noindent
 \section{Calculation results}
\subsection{An example: the $^{40}Cl+^{28}Si$ system at 40 MeV/nucleon}
 Now we discuss, as an example, the
results concerning the isospin equilibration process for the
$^{40}Cl+^{28}Si$ system at 40 MeV/nucleon.
 In Fig. 1  we show the average total dipolar signals
evaluated through CoMD-II calculations along the $\hat{z}$ beam
direction $\langle {V}^{z} \rangle$ and along the impact parameter
direction $\hat{x}$, $\langle {V}^{x} \rangle$, respectively. The
reference frame is the c.m. one. The impact parameter $b$ is equal
to 3 fm, in panels (a) and  (c) and 1.5 fm in panels (b) and (d).
In Fig. 1(a) and Fig. 1(c) the average dipolar signals are shown for
the first 140 fm/$c$.
Different lines refer to different iso-vectorial potentials,
 according
to Ref. \cite{comd1}. In the first 150 fm/$c$  in all
the cases wide oscillations exist. They are responsible for the
pre-equilibrium $\gamma$-ray emission \cite{asygdr,ca10mev}. The
damped oscillations converge towards smaller and almost constant values.
 This  can be seen in Fig. 1(b) and Fig. 1(d) in
which the dynamical evolution is followed from 110 fm/$c$ up to 300
fm/c. The inclusion of corrections due to Coulomb interaction at
longer time is  of the order of some percent.
The time interval in which the almost stationary behavior is reached
is related to the lifetime of the coherent dipolar collective mode
and it is strictly linked  with the average time for the formation of
the main fragments and pre-equilibrium emission. Fig. 2 shows the
asymptotic values of the $\hat{z}$  and
$\hat{x}$ components of the average dipolar signal, evaluated for different
reduced impact parameters $b_{r}=\frac{b}{b_{max}}$ ($b_{max}\simeq 7.5 $fm)
and different interaction options. The figure
enlightens the sensitivity of the $\langle \overrightarrow{V}
\rangle$ observable to the iso-vectorial interaction.
\vskip 5pt
\noindent
In the following we would like to discuss the behaviors of the
different components of the total dipolar signal along the $\hat{z}$
direction as a function of the impact parameter  according to the
scheme depicted in the previous section and related to Eqs. (1,4).
In Fig. 3(a) the ratio
$R_{GL}=\frac{\langle V_{G}^{z}\rangle}{\langle V_{L}^{z}\rangle}$ between the $\hat{z}$ asymptotic
components of the dipolar signal associated to the light particles $\langle V_{G}^{z}\rangle$
and the one corresponding to the two biggest fragments $\langle V_{L}^{z}\rangle$
 is plotted as a function of the reduced impact
parameter $b_{r}$  for the Stiff2
option. The other options produce similar results.
For peripheral collisions we see the predominance of the signal
carried by the two biggest fragments ($|R_{GL}|<1$). In these cases the liquid part
is dominated by the so-called "molecular" component $V_{M}$ (see
Eq.(1)). It has a negative sign because the quasi-projectile has the
largest neutron excess with respect to the quasi-target. The "gas"
component produces a small negative value which indicates, on
average, a dominance of the pre-equilibrium emission of nucleons
from the target light partner. For smaller impact parameters the
molecular component reduces its value because of the increasing
stopping (reduction of $\overrightarrow{v_{PT}}$ in Eq.(1)) and as a
consequence of the charge/mass
equilibration between the two partners. The
asymptotic behavior of $R_{GL}$ between $b_{r}=0.6-0.75$ is due to
the change of sign of the "liquid" component which evolves in a
rather continuous way passing through zero. The "liquid" part  has a positive sign
for more central collisions,
because now the process
can be roughly described as one charged source emitting
pre-equilibrium charged particles mostly on the target side. The average recoil of
the source in the projectile direction produce the positive value of
the dipolar signal.
The absolute value of $R_{GL}$ increases for more central
collisions due to the increasing contribution of the
pre-equilibrium particles.

\noindent
As discussed in the introductory section, when the collision partners are not identical
nuclei the isospin equilibration process produces a neutron-proton
differential  flow contribution if the neutron and proton "gases" have
different c.m. velocities. These pre-conditions are verified for the
studied collision.

\noindent
We briefly recall that  the neutron-proton differential flow  is an
observable that has been proposed  to investigate symmetry potential
effects \cite{bali00,bali08}. It is expressed as the average difference between the transverse
momenta of the emitted free neutrons and protons having rapidity $y'$:
\begin{equation}
F_{np}=\frac{1}{A(y')}\sum_{i=1}^{A(y')}w_{i}p_{ix}(y')
\end{equation}
\label{flow}
with $w_{i}=1$ for neutrons and $w_{i}=-1$ for protons.
For $b=3$ fm and for the Stiff1 and Soft options,
in Fig. 3(b) we show the neutron-proton differential flow $F_{np}$,
expressed in $c$ units, as a function of the particles rapidity, $y$, normalized
to the projectile one $y'_{beam}$. The rapidity values are
evaluated in the c.m. of the total system reference frame. The
dashed vertical lines indicate the projectile and target reduced
rapidity.

\noindent
From the
figure we can understand that, by averaging on the rapidity, the
neutron-proton transversal velocity has a negative value. This
reflects the average "bending" of the relative motion between the c.m. of
the emitted neutrons and protons, through the half-plane opposite to
the impact parameter direction. In Fig. 3(c) we display the average number of nucleons
as a function of the rapidity for the two options. In both the cases the collision
produces rather similar "stopping".
The results shown in Fig. 3(b) can be compared with
the calculations displayed in Fig. 3(d) obtained by subtracting,
event by event, the c.m. relative neutron-proton motion related to
the "gas" phase. As can be seen, similarly to the case of identical
nuclei, this correction restores (within the errors associated to
the statistics of simulations) the almost specular behavior of
$F_{np}$ with respect to the rapidity axes. The correction acts also
along the beam direction. The largest absolute values of $F_{np}$
around $|y|\simeq 1$ for the Soft case is mostly due to the largest neutron
excess of the "gas" phase with respect to the one obtained for the
Stiff1 option.
 Finally, in Fig. 3(e) we show the vectors associated to
the average proton-neutron relative motion
$\overrightarrow{v}_{r}^{PN}$ for the two options. These quantities
determine the correction on the differential flow due to the isospin equilibration
process and are
associated to the "gas" component of the total dipolar signal (see
first term of Eq.(4)).

\subsection{Sensitivity to the different options of the iso-vectorial
interaction}

 In the following we want to discuss in
some detail the sensitivity of the dipolar signal to different
options concerning the iso-vectorial interaction including also the
role played by the finite value of the correlation coefficient $\alpha$.

\noindent
According to the nature of the
discussed quantity,  we can expect a particular sensitivity to the
iso-vectorial forces because it contains implicitly information on
the global relative motion of the charge particles with respect the
neutral ones.
 In Fig. 4(a) we  show as a function
of the reduced impact parameter $b_{r}$
 the asymptotic values of $\langle {V}^{x}
\rangle$ and $\langle {V}^{z} \rangle$. The arrow indicates the
direction of increasing impact parameters corresponding to the
marked points with a step $\Delta b_{r}=0.1$. Different symbols
represent different options. In the region of the bending of the
lines, around $b=3$ fm, we observe the greater sensitivity to the
iso-vectorial interactions. This result is particularly
evident by studying the ratios $R=\langle {V}^{x} \rangle / \langle
{V}^{z} \rangle$.
In Fig. 4(b) we in fact show the relative change
       $r=\frac{\Delta R}{R}$
 between couples of different options. We can see that for
$b_{r}$ less than about 0.6 large changes are predicted according to
the different shapes of the form factor $F'$. This impact parameter region produces
a substantial stopping of the incident nuclei and  is
clearly dominated by large overlap between projectile and target
 which gives rise to processes changing from incomplete fusion
reactions to IMF  production . The region of intermediate impact
parameters show the higher sensitivity when the mechanism evolves
with respect to the essentially binary processes which take place at
the higher impact parameters. For $b_{r}$ greater than 0.6, in fact,
the sensitivity is strongly reduced.

\noindent
In particular, according to what previously observed  (see for
example point (ii) of Sec. I), we have evaluated the partial contributions
$\langle {V}^{x}_{L} \rangle $ and $\langle {V}^{z}_{L} \rangle$ related to the two
main fragments. As an example, for $b=3$ fm and for the Stiff2
option,  the "liquid" asymptotic values are
$\langle {V}^{x}_{L} \rangle=-0.120 c  $ and $\langle {V}^{z}_{L} \rangle=0.162 c $
while the total contributions are $\langle {V}^{x} \rangle=0.044 c$ and
$\langle {V}^{z} \rangle=-0.027 c $. Therefore, it results  that the
contributions carried by the two main fragments only partially
contribute to the isospin equilibration process. The remaining part
("gas"), which in this case we have associated to
 particles and to the
IMF, generates a term with opposite sign and similar strength for
both directions. It contributes in a decisive way to the global
equilibration process. For the same impact parameter,
in Fig. 4(b) we show with the star symbol the sensitivity parameter
$r$ evaluated by changing the option from Stiff1 to Stiff2
 and by only taking into account the contributions of the two main fragments.
As we can see, the partial contribution shows  a
rather reduced sensitivity to the different options as compared to
the case  obtained by using the global information on the system.

\vskip 5pt \noindent
 Finally, in the following we show the role of the
correlation coefficient $\alpha$, introduced in Sec.II, into
determine the sensitivity of the investigated observable to the
density dependence of the iso-vectorial interaction. For this aim,
in Fig. 4(c), we compare, for different impact parameters, the values
of $r$ obtained for the Stiff2-Stiff1 options with the ones obtained
in the I.M.F.A. case.
  Fig. 4(c) clearly shows that in
the I.M.F.A. case, at the investigated energies, the sensitivity of
the isospin equilibration process to the behavior of the symmetry
interaction is rather reduced.  The I.M.F.A. also strongly affects
the values of $\langle \overrightarrow{V} \rangle$. In particular,
independently from the used options, it produces for $b_{r}$=0.4,
values of $|\langle {V}^{z} \rangle|$ about four times larger than  the
ones obtained with full CoMD-II calculations indicating a reduced
capacity to obtain isospin equilibration along the $\hat{z}$
direction.

\section{Summary and conclusive remarks}

In summary, in this work the isospin equilibration process
has been investigated by studying  the ensemble average of the time
derivative of the total dipole $\langle \overrightarrow{V} \rangle$
evaluated through CoMD-II calculations. Some general properties of
this quantity have been discussed. In particular, it allows to
generalize the definition of isospin equilibration also in complex
reactions evolving through multi-fragmentation processes. As an example,
calculations performed for the
asymmetric charge/mass system $^{40}Cl+^{28}Si$ at 40 MeV/nucleon
show that the  asymptotic values of
$\langle \overrightarrow{V} \rangle$ for these processes are quite
sensitive to different options for the iso-vectorial potential; moreover, in
central and mid-peripheral collisions, the dipolar contribution associated to
the pre-equilibrium emission of charged particles is relevant  to
determine the value of $\langle \overrightarrow{V} \rangle$ and the
related sensitivity to different density dependent form factors.
Semiclassical CoMD-II calculations performed in the so-called I.M.F.A. scheme also
highlights  the prominent role which could be played by the two-body
neutron-proton correlations in the study of the dynamics leading to the
isospin equilibration processes.

\vskip 10pt
\eject \noindent

Fig. 1 - Average dipolar signals $\langle \overrightarrow{V} \rangle$ for
$b$=3 fm along the $\hat{z}$ direction are plotted as a function of
time in the intervals  0-140 fm/$c$ (panel (a)) and 110-300 fm/$c$
(panel (b)). Different lines indicate different options for the
iso-vectorial interaction (see the text). Panels (c) and (d) display the
average dipolar signals  along the $\hat{x}$ direction for $b$=1.5
fm and in the same time intervals like panels (a) and (b)
respectively.

\vskip5pt \noindent

Fig. 2 - Asymptotic values $\langle {V}^{x} \rangle$ and $\langle {V}^{z} \rangle$
evaluated for different values of
 the $b_{r}$ parameter (see the text). Different symbols indicate different
 options for the iso-vectorial interaction.

\vskip5pt
\noindent

 Fig. 3 - In panel (a) the $R_{GL}=\frac{\langle V_{G}^{z}\rangle}{\langle V_{L}^{z}\rangle}$
 ratio (see the
text) is plotted as a function of the reduced impact parameter for the Stiff2 option.
Panels (b) and (c) display the neutron-proton differential flow $F_{np}$ and the
number of free nucleons $A_{G}$ as a
function of the c.m. rapidity respectively.
In the the panel (d), $F_{np}$ is plotted after the
correction for the relative c.m. velocity  of the
neutron and proton "gases" $\overrightarrow{v}_{r}^{PN}$. The vertical dotted lines
 represent the target and projectile rapidity.
 Different symbols refers to different options.
In the panel (e)  the average  $\langle \overrightarrow{v}_{r}^{PN} \rangle$ contributing
to the isospin equilibration process is represented for the Stiff1 and Soft options.
\vskip5pt \noindent

 Fig. 4 -(a)
 $\langle {V}^{x} \rangle$ is plotted as a
 function of the corresponding $\langle {V}^{z} \rangle$
 value for different reduced impact parameter $b_{r}$ values
 ($b_{max}\simeq 7.5 $ fm and
$\Delta b_{r}=0.1$ ) and different options
 (different symbols). The arrow indicates the direction
 of increasing impact parameters.
 -(b) relative changes $r$ for the ratio $R$ (see the text)
 evaluated for different couples of options as a function of
 $b_{r}$. -(c) Values of $r$ evaluated
for the Stiff1-Stiff2 options as a function of $b_{r}$ are plotted
in the case of full CoMD-II calculations and in the case of
I.M.F.A. approximation (see the text). The lines which join the
points are meant only to guide the eye through the shown trend.

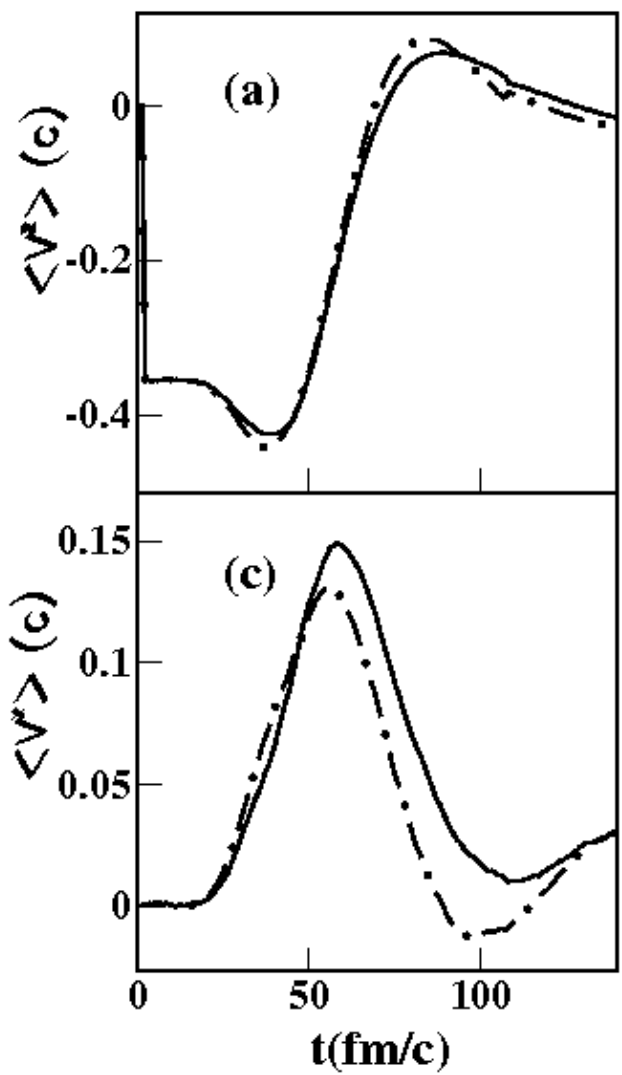
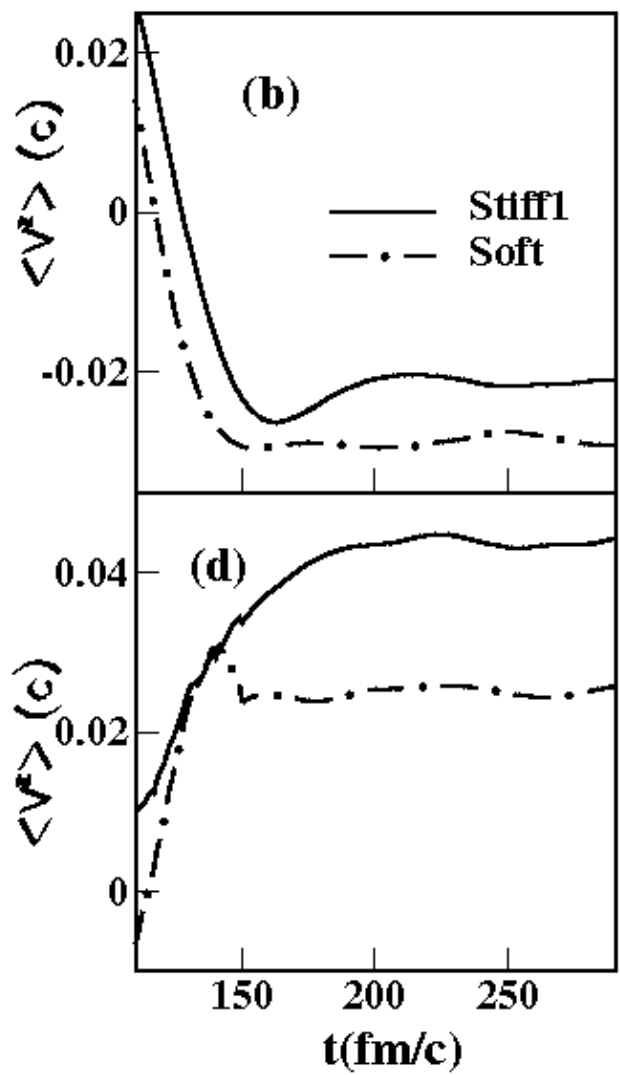
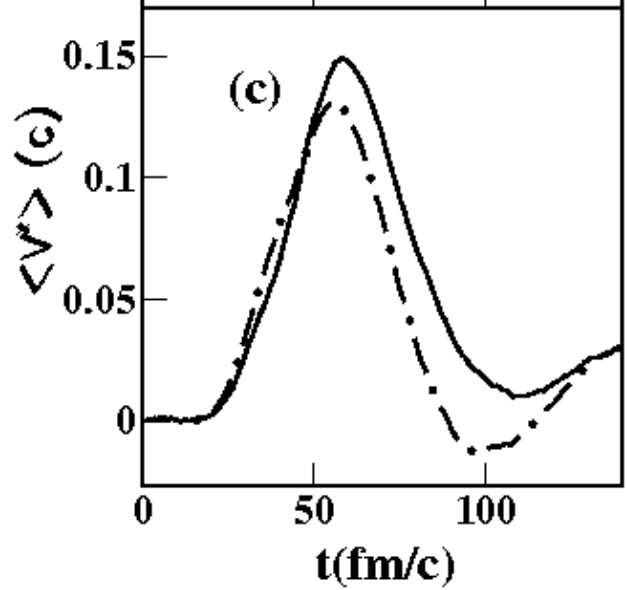
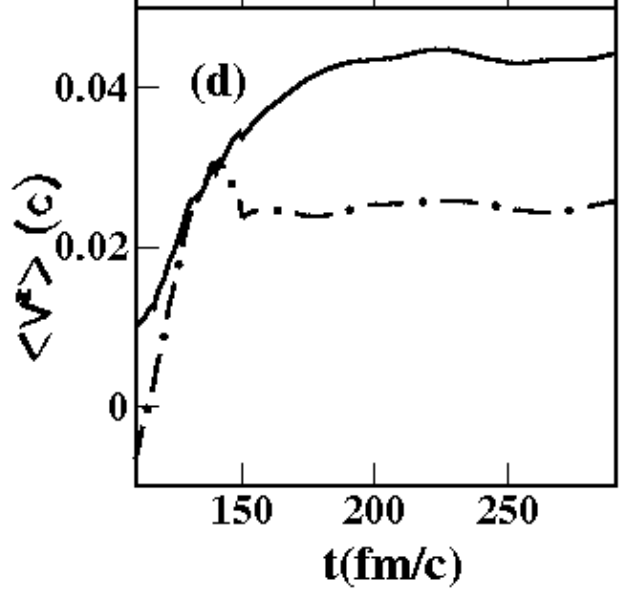
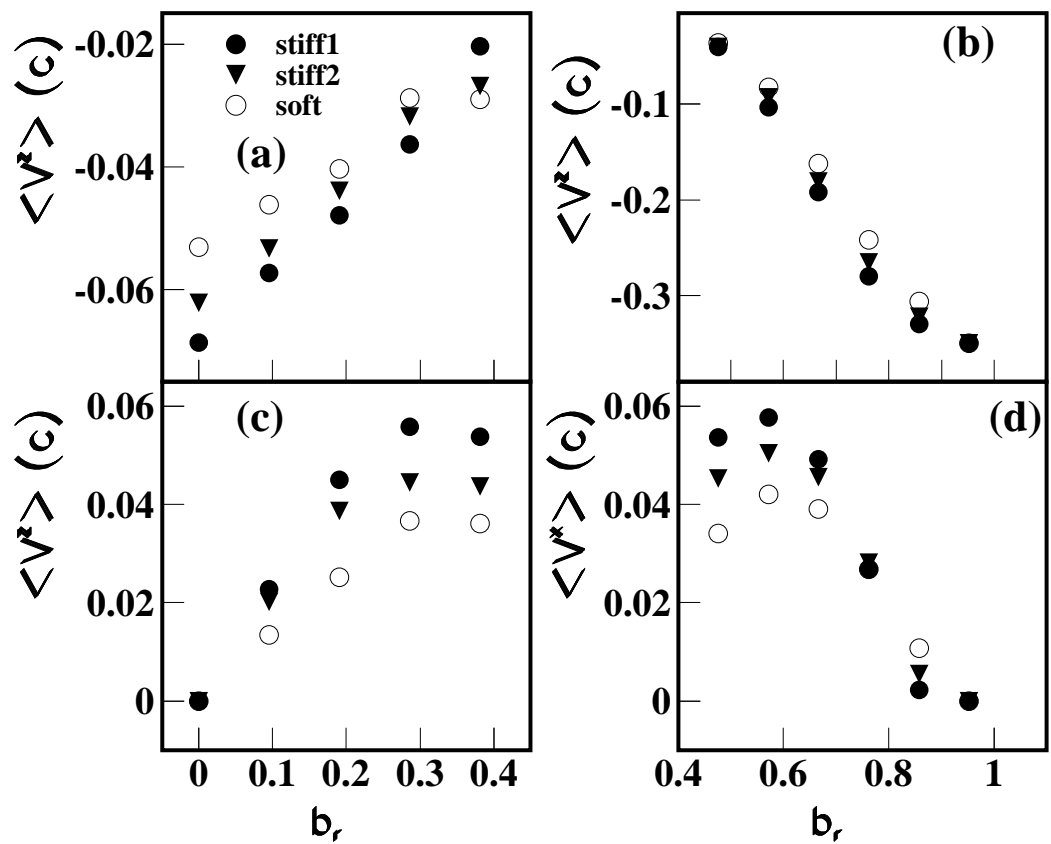
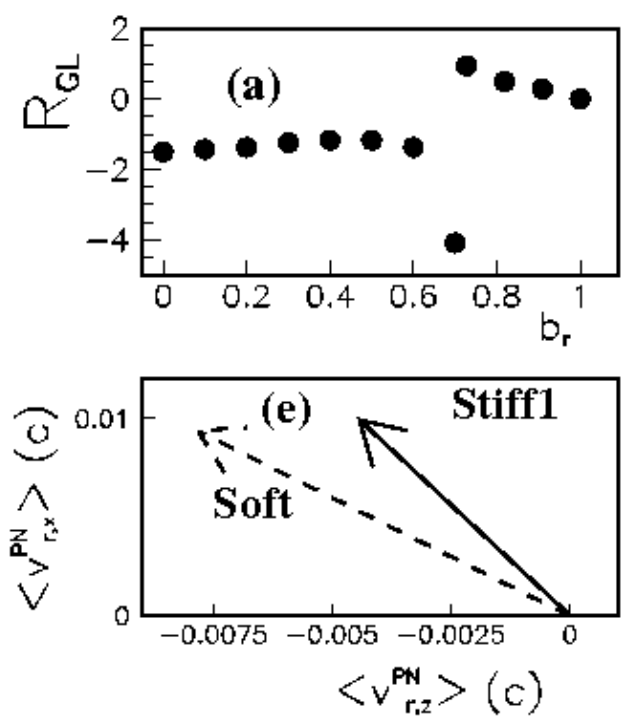
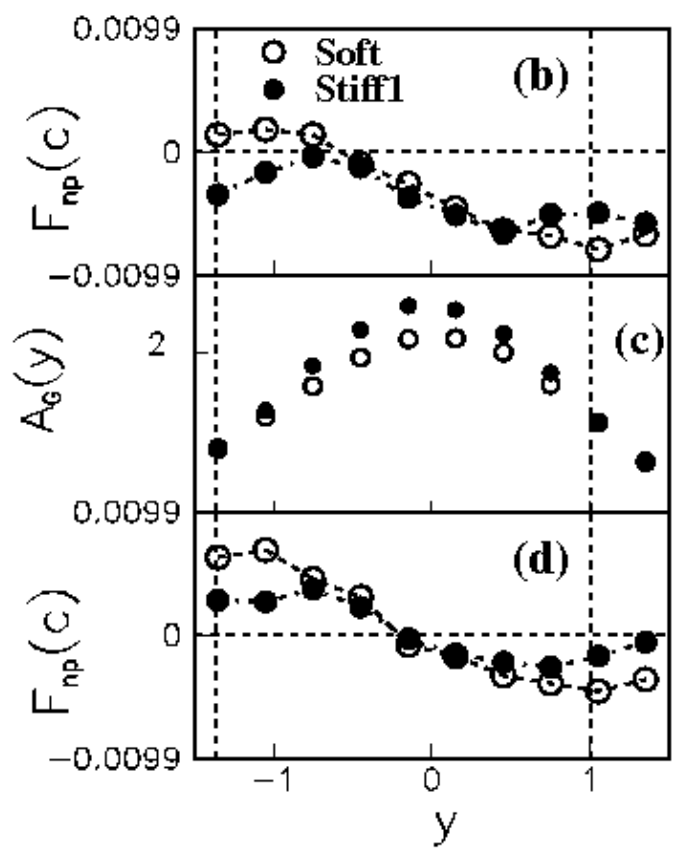
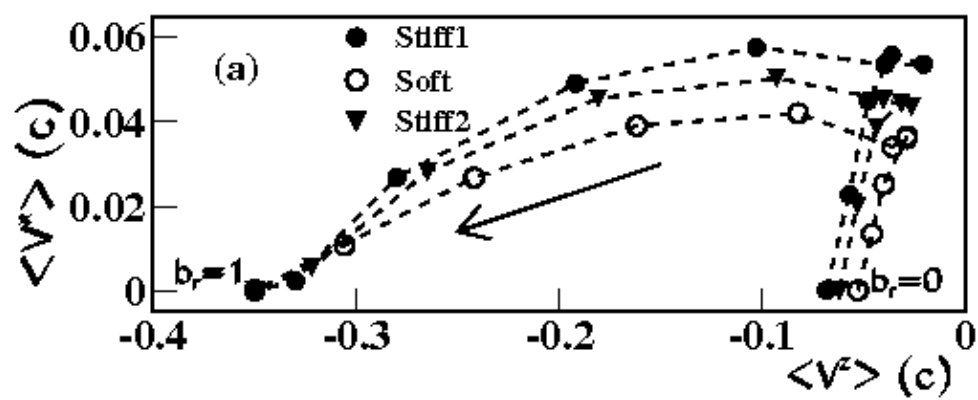
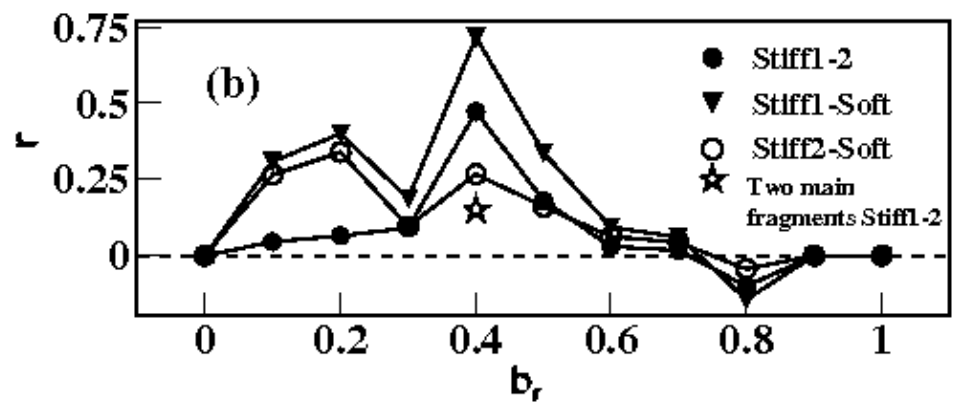
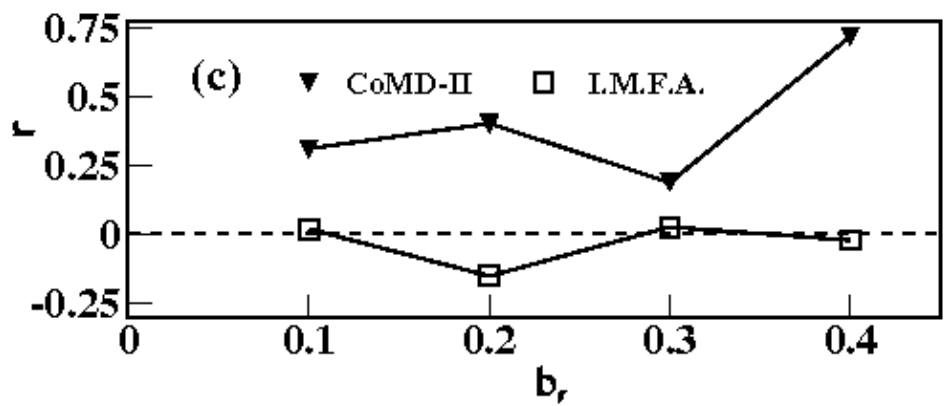
\end{document}